\documentclass[iop]{emulateapj}
\usepackage{amsmath}
\usepackage{multirow}

\shorttitle{Isochrone Comparison using Multi-Band Photometry}
\shortauthors{Thompson et al.}

\begin{document}

\title{WIYN Open Cluster Study LXII: Comparison of Isochrone Systems using Deep Multi-Band Photometry of M35}

\author{B. Thompson\altaffilmark{1}}
\author{P. Frinchaboy\altaffilmark{1}}

\author{K. Kinemuchi\altaffilmark{2}}

\author{A. Sarajedini\altaffilmark{3}}

\author{R. Cohen\altaffilmark{4}}

\altaffiltext{1}{Texas Christian University, Fort Worth, TX 76129}
\altaffiltext{2}{Apache Point Observatory, Sunspot, NM 88349}
\altaffiltext{3}{University of Florida, Gainesville, FL 32611}
\altaffiltext{4}{Universidad de Concepcion, Concepcion, Chile}

\begin{abstract}
Current generation stellar isochrone models exhibit non-negligible discrepancies due to variations in the input physics. The success of each model is determined by how well it fits the observations, and this paper aims to disentangle contributions from the various physical inputs. New deep, wide-field optical and near-infrared photometry ($UBVRIJHK_S$) of the cluster M35 is presented, against which several isochrone systems are compared: Padova, PARSEC, Dartmouth and Y$^2$. Two different atmosphere models are applied to each isochrone: ATLAS9 and BT-Settl. For any isochrone set and atmosphere model, observed data are accurately reproduced for all stars more massive then $0.7$ M$_\odot$. For stars less massive than 0.7 M$_\odot$, Padova and PARSEC isochrones consistently produce higher temperatures than observed. Dartmouth and Y$^2$ isochrones with BT-Settl atmospheres reproduce optical data accurately, however they appear too blue in IR colors. It is speculated that molecular contributions to stellar spectra in the near-infrared may not be fully explored, and that future study may reconcile these differences.
\end{abstract}


\section{Introduction}
Aside from white dwarf cooling, the main sequence is perhaps the most well-understood part of stellar evolution. Yet, current generation stellar structure models show significant discrepancies along parts of the main sequence due to adoption of different input physics. The usefulness of a model depends on how well it fits data, the most common method being comparing stellar isochrones to observed star cluster color-magnitude diagrams (CMDs). While the common method of comparing results from different models to observed data gives an excellent first estimate, the aim of this work is to disentangle contributions from various physical inputs into the models. Ideally, the tests would compare multiple different stellar isochrones to a series of open clusters, allowing a determination of which underlying physical parameters lead to an accurate fit.

This process is started by testing isochrone models against the open cluster M35. M35 provides a good starting point for the analysis due to its young age of 178 Myr \citep{2002AA...389..871D}, as all stars in the cluster have had time to settle onto the main sequence, but not enough time for many to evolve off of it. M35 will allow probing of how well various isochrone models work for main sequence stars over a large mass range, from 0.3 -- 3.0 M$_\odot$.

\section{Observational Data} \label{sec:data}
To provide a comprehensive test of the isochrone models to M35, accurate, multi-wavelength photometry is required. We present new photometry for M35 in both the optical and infrared, used in this analysis.

\begin{figure*} \centering
\includegraphics[trim = 0mm 20mm 20mm 0mm, clip, width=3.3in]{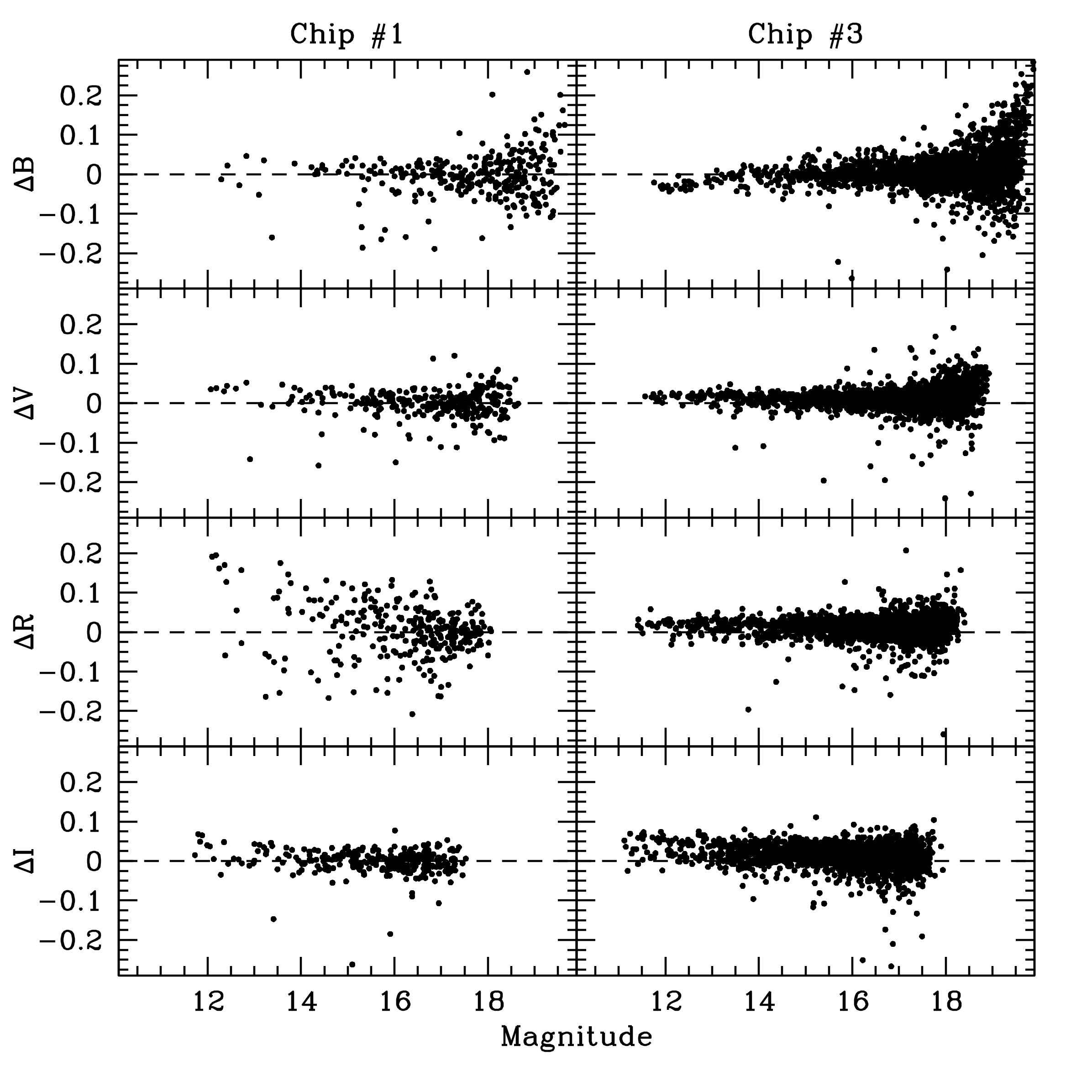}			
\includegraphics[trim = 0mm 20mm 20mm 40mm, clip, width=3.3in]{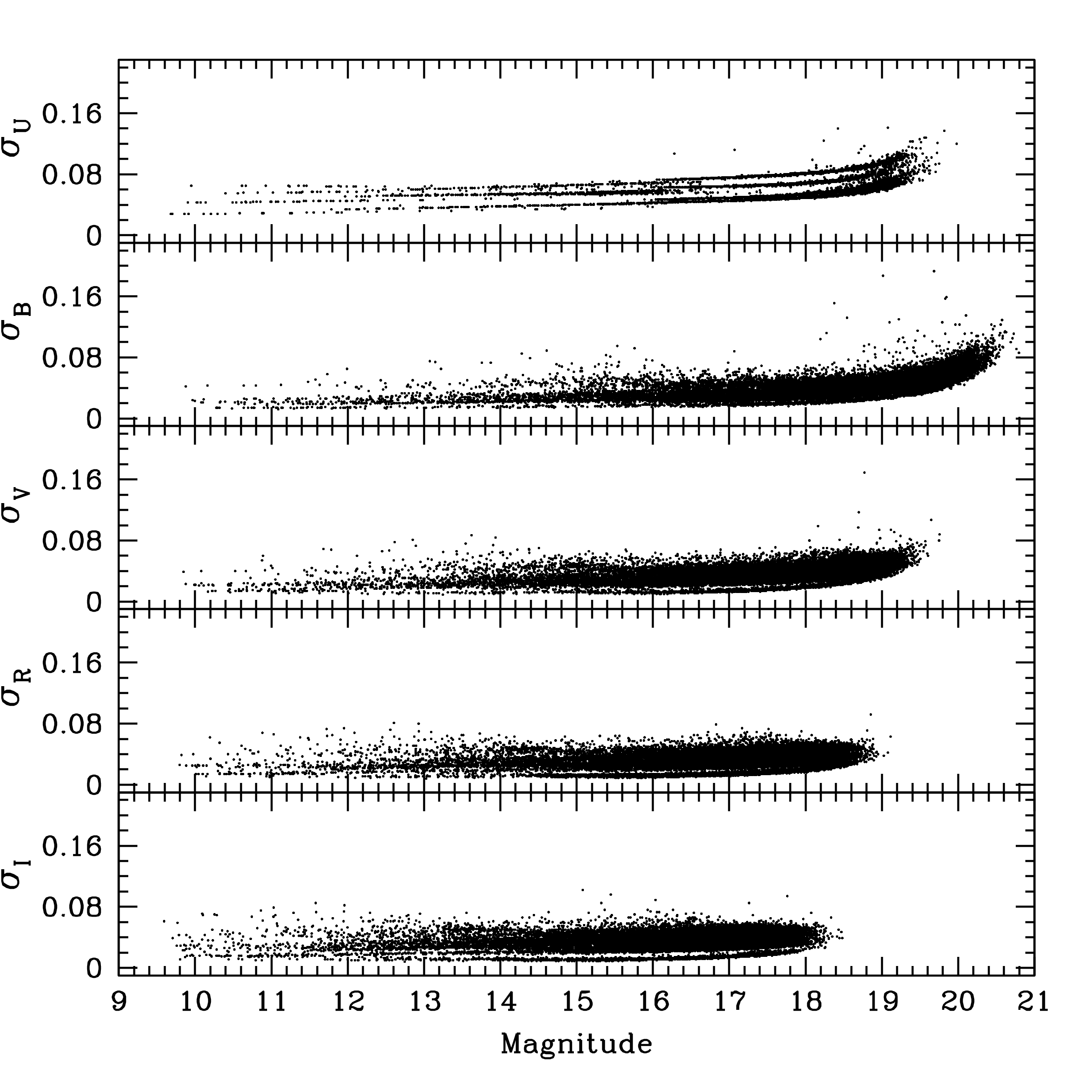}		
\caption{\emph{Left:} Residuals for $BVRI$ in calibrating the MOSAIC data. Chip \#1 $R$--band data contains a secondary color transformation to improve the errors in the fit. \emph{Right:} MOSAIC Magnitude vs uncertainty for the combined dataset. Differing transformation errors between chips are visible for each filter. \label{fig:MOSphotometry}}
\end{figure*}

\begin{figure*} \centering
\includegraphics[trim = 0mm 20mm 20mm 200mm, clip, width=3.3in]{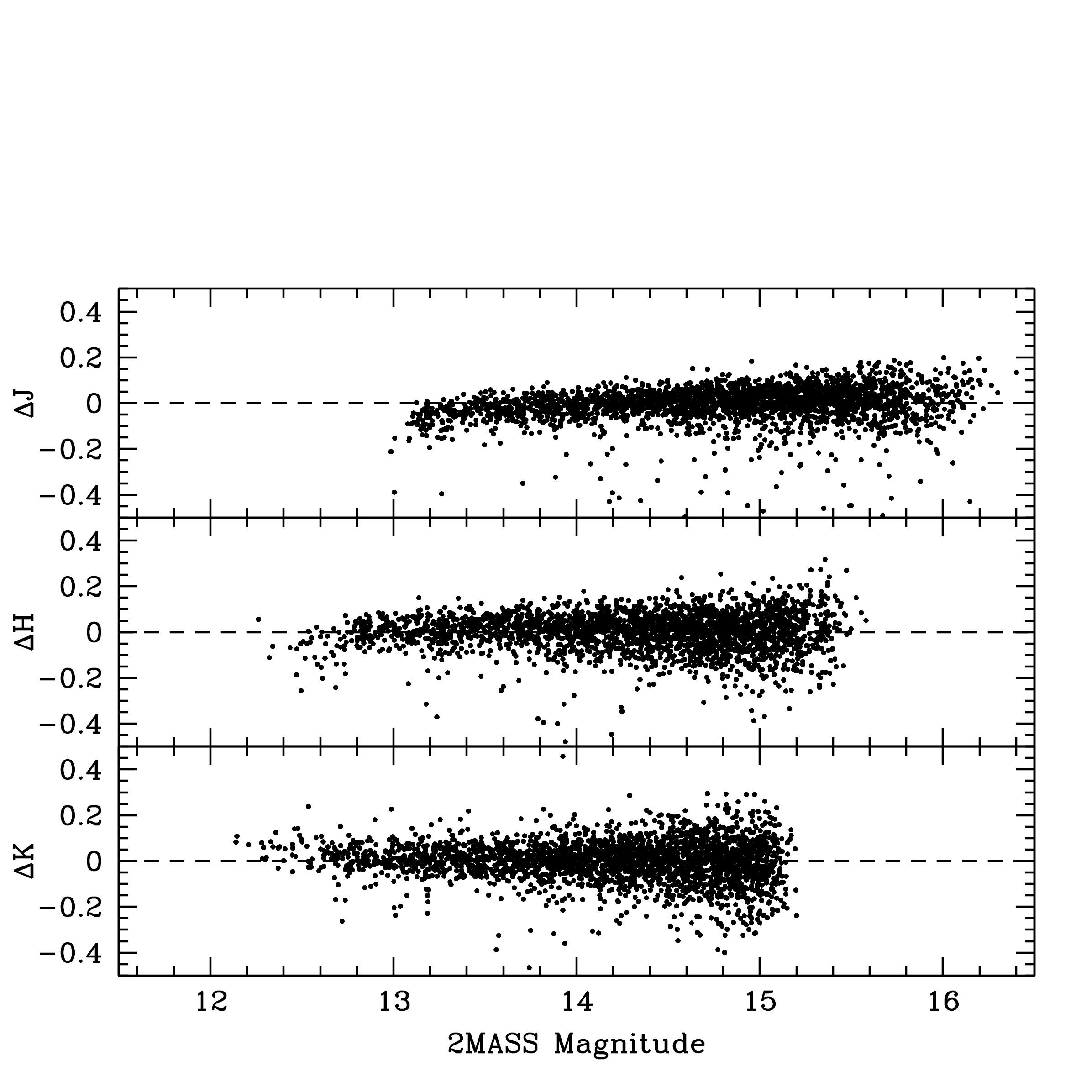}		
\includegraphics[trim = 0mm 20mm 20mm 200mm, clip, width=3.3in]{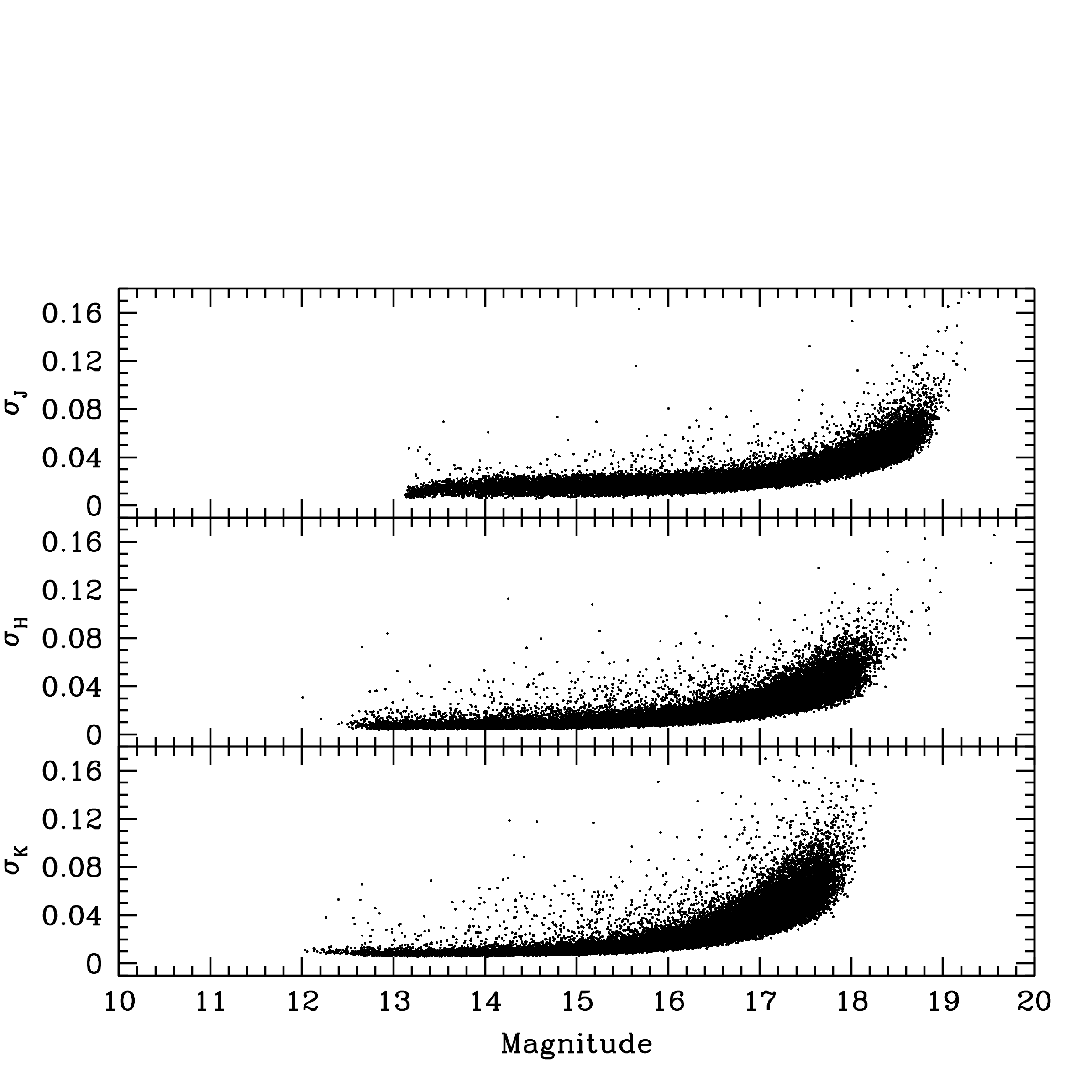}		
\caption{\emph{Left:} Residuals between instrumental NEWFIRM magnitudes and 2MASS. \emph{Right:} NEWFIRM Magnitude vs uncertainty for all four pointings on M35.\label{fig:NFMphotometry}}
\end{figure*}

\subsection{$UBVRI$ Photometry}
M35 is a well-studied cluster, having been observed in the optical many times before. The first CCD photometry of the cluster was published by \citet{1999MNRAS.306..361S}, and more recently several other studies \citep{2002AJ....124.1555V, 2003AJ....126.1402K, 2000AAS...196.4206S} have published deep photometry on the cluster. Previous WIYN Open Cluster Survey (WOCS) photometry work \citep{2002AJ....124.1555V} published $BVI$ on a $20.5^\prime \times 20.5^\prime$ field of view around the cluster, while \citet{2003AJ....126.1402K}, using the CFH12K mosaic camera \citep{2001ASPC..232..398C}, published $BV$ photometry on a $42^\prime \times 28^\prime$ area.

Using the KPNO 0.9-m MOSAIC camera \citep{2010SPIE.7735E.111S}, a $59^\prime \times 59^\prime$ field of view has been observed in $UBVRI$, increasing both the spatial and wavelength coverage beyond previous studies.

M35 images, observed over two nights in February 2000, were taken in two sequences: short and long, allowing for photometry of both the brightest and faintest stars in the cluster. Short exposures consisted of four images per filter with exposure lengths of 25s, 8s, 5s, 3s, 5s in $UBVRI$, respectively. Long exposures were also sets of four images per filter, but 10 times the exposure length of the shorter set: 250s, 80s, 50s, 30s, 50s.

Photometry was completed using the DAOPHOT II and ALLSTAR programs \citep{1987PASP...99..191S}. A detection threshold of $3\sigma$ was used, and initially 1,000 stars were chosen to compute a point-spread function (PSF) for the frame. Stars were removed from the PSF list that fell within 4 full width at half-maximum (FWHM) from another detected source, ensuring the PSF was not contaminated by crowded stars. Next, stars that were near bad or saturated pixels were removed. Lastly, stars whose PSF $\chi^2$ fit values were more than $2\sigma$ above the mean were removed. After these removals, $400-600$ ``clean'' stars remained, from which a PSF was determined. The PSF was allowed to vary quadratically across the frame.

The instrumental magnitudes from DAOPHOT were matched to previously calibrated $UBVRI$ observations of M35 \citep{2000AAS...196.4206S}. Using between 200 and 1000 stars, depending upon filter, transformation equations were determined of the form:
\begin{equation} \label{eq:Utrans}
	u = U + a_U + b_U \times (U-B)
\end{equation}
\begin{equation} \label{eq:Btrans}
	b = B + a_B + b_B \times (B-V)
\end{equation}
\begin{equation} \label{eq:Vtrans}
	v = V + a_V + b_V \times (B-V)
\end{equation}
\begin{equation} \label{eq:Rtrans}
	r = R + a_R + b_R \times (V-R)
\end{equation}
\begin{equation} \label{eq:Itrans}
	i = I + a_I + b_I \times (R-I)
\end{equation}

Here, lower case letters represent instrumental magnitudes, while uppercase letters represent calibration magnitudes. The MOSAIC instrument contains a 8k $\times$ 8k pixel camera, comprised of eight 2k $\times$ 4k detector chips. Transformation coefficients in these equations were found to vary between each of the 8 chips. Transformation coefficients for all filters and chips are listed in table \ref{tab:MOScoeffs}. Calibrating photometry in $U$ was only available for the middle four chips (chips 2, 3, 6, \& 7). $U$-band photometry in the outlying chips was removed from the dataset. 

In constructing the transformation, equations for calibrating colors were also derived. In doing so, it was discovered that for most chips there were no noticeable trends due to color. In chip 1, however, there appeared to be an effect related to the $R$-band filter, where a large scatter was observed in the residuals, as seen in figure \ref{fig:MOSphotometry}. This scatter also appeared in the comparison of the $(R-I)$ color term, and was partially corrected out with another transformation equation:
\begin{equation}
	(r-i) = a_{RI} + b_{RI} \times (R-I)
\end{equation} 

This color correction was only applied to the long exposure set of chip 1, as there was much less of an effect in the shorter exposures.

\begin{table*} \centering 
\caption{Transformation coefficients for MOSAIC calibration \label{tab:MOScoeffs}}
\begin{tabular}{c|c|c|c|c|c} \hline \hline
Chip 				& 		U 					& 			B 				& 		V 					& 		R 					& 		I 			 \\ \hline
\multirow{2}{*}{1} 	& ...						& $a_B = 1.275 \pm 0.032$ & $a_V = 0.627 \pm 0.032$  & $a_R = 0.530 \pm 0.029$ 	& $a_I = 0.529 \pm 0.027$ \\
					& ...						& $b_B = 0.091 \pm 0.035$ & $b_V = -0.079 \pm 0.035$ & $b_R = -0.385 \pm 0.052$	& $b_I = 0.086 \pm 0.046$ \\ \hline

\multirow{2}{*}{2} 	& $a_U = 0.537 \pm 0.010$ & $a_B = 1.203 \pm 0.020$ & $a_V = 0.555 \pm 0.018$ 	& $a_R = 0.263 \pm 0.018$ 	& $a_I = 0.442 \pm 0.019$ \\
					& $b_U = 0.131 \pm 0.025$ & $b_B = 0.169 \pm 0.022$	& $b_V = -0.009 \pm 0.020$	& $b_R = 0.080 \pm 0.033$	& $b_I = 0.066 \pm 0.034$ \\ \hline
					
\multirow{2}{*}{3} 	& $a_U = 0.496 \pm 0.010$ & $a_B = 1.207 \pm 0.020$ & $a_V = 0.588 \pm 0.020$ 	& $a_R = 0.299 \pm 0.021$ 	& $a_I = 0.496 \pm 0.023$ \\
					& $b_U = 0.131 \pm 0.025$ & $b_B = 0.160 \pm 0.022$	& $b_V = -0.052 \pm 0.021$	& $b_R = 0.008 \pm 0.038$	& $b_I = -0.039 \pm 0.042$ \\ \hline
					
\multirow{2}{*}{4} 	& ...					  & $a_B = 1.133 \pm 0.026$ & $a_V = 0.506 \pm 0.026$ 	& $a_R = 0.220 \pm 0.027$ 	& $a_I = 0.387 \pm 0.030$ \\
					& ...					  & $b_B = 0.202 \pm 0.028$	& $b_V = -0.003 \pm 0.029$	& $b_R = 0.082 \pm 0.049$	& $b_I = 0.060 \pm 0.053$ \\ \hline
					
\multirow{2}{*}{5} 	& ...					  & $a_B = 1.215 \pm 0.013$ & $a_V = 0.581 \pm 0.008$ 	& $a_R = 0.297 \pm 0.007$ 	& $a_I = 0.476 \pm 0.007$ \\
					& ...					  & $b_B = 0.120 \pm 0.014$	& $b_V = -0.048 \pm 0.009$	& $b_R = 0.027 \pm 0.012$	& $b_I = -0.025 \pm 0.011$ \\ \hline
					
\multirow{2}{*}{6} 	& $a_U = 0.527 \pm 0.016$ & $a_B = 1.176 \pm 0.018$ & $a_V = 0.545 \pm 0.016$ 	& $a_R = 0.235 \pm 0.015$ 	& $a_I = 0.466 \pm 0.016$ \\
					& $b_U = 0.190 \pm 0.036$ & $b_B = 0.202 \pm 0.019$	& $b_V = -0.010 \pm 0.017$	& $b_R = 0.084 \pm 0.027$	& $b_I = -0.003 \pm 0.027$ \\ \hline
					
\multirow{2}{*}{7} 	& $a_U = 0.481 \pm 0.031$ & $a_B = 1.170 \pm 0.023$ & $a_V = 0.555 \pm 0.022$ 	& $a_R = 0.227 \pm 0.024$ 	& $a_I = 0.417 \pm 0.025$ \\
					& $b_U = 0.194 \pm 0.077$ & $b_B = 0.200 \pm 0.024$	& $b_V = -0.026 \pm 0.023$	& $b_R = 0.080 \pm 0.042$	& $b_I = 0.042 \pm 0.045$ \\ \hline
					
\multirow{2}{*}{8} 	& ...					  & $a_B = 1.170 \pm 0.035$ & $a_V = 0.561 \pm 0.031$ 	& $a_R = 0.246 \pm 0.032$ 	& $a_I = 0.481 \pm 0.034$ \\
					& ...					  & $b_B = 0.165 \pm 0.036$	& $b_V = -0.063 \pm 0.033$	& $b_R = -0.012 \pm 0.056$	& $b_I = -0.127 \pm 0.058$ \\ \hline
\end{tabular}
\end{table*}

After transformation, the photometry was combined for all images. Stars detected in multiple images had their magnitudes combined via an error-weighted average.

\subsection{$JHK_S$ Near-IR Photometry}
2-Micron All Sky Survey \citep[2MASS;][]{2006AJ....131.1163S} near-IR photometry is available over the entire sky, providing $JHK_S$ magnitudes for stars in M35. Isochrone comparisons using deep optical and 2MASS data were published in previous WOCS work \citep{2003MNRAS.345.1015G}. While providing insight on differences between isochrone systems in optical bands, IR comparisons were limited to fairly bright stars. Low-mass members of the cluster are important in the cluster's dynamical evolution, and accurately determining their parameters is critical for stellar structure models. To fully compare isochrones in the IR, deeper photometry was necessary.

Observations of M35 were taken using the NEWFIRM instrument \citep{2004SPIE.5499...59H} on the Kitt Peak 4-m telescope in February 2008. The NEWFIRM camera is a grid of four 2k $\times$ 2k IR detectors, creating a 4k $\times$ 4k image. All observations were taken in ``4Q'' mode, aligning the cluster within each of these four NEWFIRM detectors, allowing for more spatial coverage than a single NEWFIRM field of view. Together, the images cover a $44^\prime \times 44^\prime$ area around the cluster. To minimize errors in flat-fielding and negate cosmetic defects within the detectors, the telescope was dithered between exposures on each pointing. An effective integration time of 600 seconds in $J$ and $H$, and 900 seconds in $K_S$ were taken for each pointing.

All images were reduced through the NEWFIRM Pipeline \citep{2009ASPC..411..506S}. After reduction, images were stacked into a master frame for each filter. Photometry on these frames were carried out using DAOPHOT II and ALLSTAR. Initially, 2000 stars were chosen to determine a PSF for the frame, and the list was trimmed using the same process as the MOSAIC data: crowded stars (less than 4 FWHM from another source), saturated stars, stars near bad pixels, and those with $\chi^2$ values more than $2\sigma$ above the mean were removed. After cleaning, between 700 and 900 uncrowded stars per frame were used to compute a PSF, which was allowed to vary quadratically.

2MASS data were used to tie the instrumental magnitudes to the standard system. Only 2MASS point sources with the highest photometric quality (`AAA') were used in the reference catalog. Matching more than 2,500 2MASS stars in each filter to the DAOPHOT instrumental magnitudes, transformations to the standard system were determined as:
\begin{equation} \label{eq:Jtrans}
	j = J - (2.400 \pm 0.003) - (0.0987 \pm 0.005) \times ( J-K_S)
\end{equation}
\begin{equation} \label{eq:Htrans}
	h = H - (2.297 \pm 0.002) - (0.2956 \pm 0.012) \times ( H-K_S)
\end{equation}
\begin{equation} \label{eq:Ktrans}
	k = K_S - (3.030 \pm 0.005) + (0.093 \pm 0.007) \times ( J-K_S)
\end{equation}

As before, lower case letters are instrumental magnitudes, while upper case letters are standard 2MASS magnitudes. A plot of residuals from this transformation is shown in figure \ref{fig:NFMphotometry}.

\subsection{Merged Dataset}
\begin{figure*} \centering
\includegraphics[trim = 0mm 28mm 20mm 20mm, clip, width=6in]{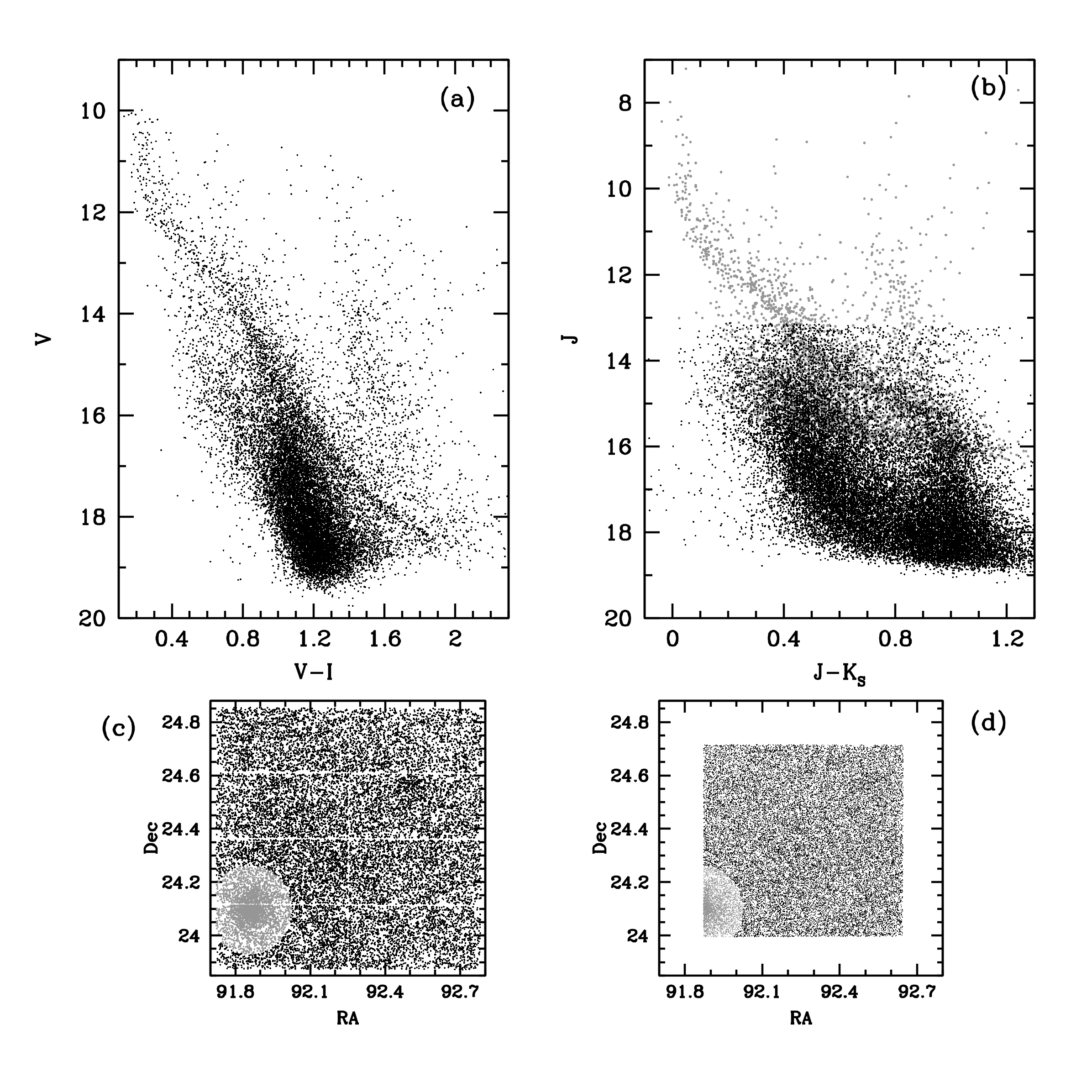}		
\caption{CMDs and spatial plots for each of the new datasets. \emph{(a)} Visual CMD of MOSAIC dataset. \emph{(b)} Near-IR CMD of NEWFIRM dataset. Grey points are `AAA'-quality 2MASS magnitudes. \emph{(c)} Spatial plots of MOSAIC detections. A small area around NGC 2158 has been removed, denoted by grey points. \emph{(d)} Spatial plot of NEWFIRM detections. Removed area around NGC 2158 denoted by grey points. M35 is located in the center of the spatial plots. \label{fig:finalphot}}
\end{figure*}

With 600s or more of exposure time on a 4-m telescope, stars with $J < 13$ are saturated in the NEWFIRM images. In the final dataset, the MOSAIC and NEWFIRM photometry was merged with all `AAA'-quality 2MASS point sources to form a complete picture of the cluster in the optical and near-IR. The merged dataset is shown in table \ref{tab:stubtable}; CMDs and spatial diagrams for this dataset are shown in figure \ref{fig:finalphot}.

\begin{table*}\tiny
\caption{Combined Dataset Photometry \label{tab:stubtable}}
\begin{tabular}{cccccccccccccccccc}
RA & Dec & $U$ & $\sigma\_U$ & $B$ & $\sigma\_B$ & $V$ & $\sigma\_V$ & $R$ & $\sigma\_R$ & $I$ & $\sigma\_I$ & $J$ & $\sigma\_J$ & $H$ & $\sigma\_H$ & $K_S$ & $\sigma\_K_S$ \\ \hline
   91.88346  &   24.43455 & 99.999 &  9.999 & 19.878 &  0.045 & 18.873 &  0.037 & 18.268 &  0.033 & 17.688 &  0.032 & 16.744 &  0.018 & 16.425 &  0.021 & 16.134 &  0.021 \\
   92.34281  &   24.43458 & 99.999 &  9.999 & 19.543 &  0.045 & 18.669 &  0.034 & 18.051 &  0.028 & 17.446 &  0.026 & 16.555 &  0.018 & 16.100 &  0.010 & 15.904 &  0.012 \\
   92.56734  &   24.43462 & 99.999 &  9.999 & 19.590 &  0.041 & 18.605 &  0.033 & 18.027 &  0.027 & 17.377 &  0.027 & 16.517 &  0.016 & 16.073 &  0.010 & 15.909 &  0.014 \\
   92.43966  &   24.43469 & 16.700 &  0.075 & 16.422 &  0.022 & 15.845 &  0.019 & 15.503 &  0.018 & 15.099 &  0.020 & 14.490 &  0.017 & 14.275 &  0.010 & 14.084 &  0.007 \\
   92.30543  &   24.43478 & 18.854 &  0.093 & 18.678 &  0.030 & 17.831 &  0.025 & 17.273 &  0.023 & 16.715 &  0.024 & 15.953 &  0.021 & 15.585 &  0.007 & 15.443 &  0.009 \\ \hline
\end{tabular}
{\scriptsize This table is available in its entirety in machine-readable form in the online journal. A portion is shown here for guidance regarding its form and content.}
\end{table*}

All images of M35 also cover the nearby cluster, NGC 2158. To reduce contamination, stars within close proximity (10$^\prime$) to NGC 2158 were removed. Even with the trimming of NGC 2158, a large amount of field contamination still remains due to M35's low galactic latitude. To limit this contamination, CMDs of M35 analyzed in this paper will be limited to stars within 20$^\prime$ of the cluster center.

\section{Stellar Structure Models} \label{sec:isochrones}

\begin{table*} \centering \tiny
\caption{Differences in input physics between stellar structure models used in this work. \label{tab:isoparams}}
\begin{tabular}{c||c|c|c|c} \hline \hline
	& Dartmouth & Y$^2$ & Padova & PARSEC \\ \hline
	\multirow{2}{*}{Opacity} & $\log T > 4.5$: OPAL96\footnote{ \citet{1996ApJ...464..943I}} & $\log T > 4.1$: OPAL96$^\text{a}$ & $\log T > 4.1$: OPAL93\footnote{\citet{1993ApJ...412..752I}} & $\log T > 4.2$: OPAL96$^\text{a}$ \\
			& $\log T < 4.3$: \citet{2005ApJ...623..585F} & $\log T < 3.8$: AF94\footnote{\citet{1994ApJ...437..879A}} & $\log T < 4.0$: AF94$^c$ & $\log T < 4.1$: AESOPUS\footnote{ \citet{2009AA...508.1539M}} \\ \hline
	\multirow{2}{*}{Eq Of State} & $M > 0.8 M_\odot$: Ideal Gas + Debye Huckel & \multirow{2}{*}{OPAL\footnote{\citet{1996ApJ...456..902R}}} & $\log T > 7$: \citet{1965ZA.....61..241K} & \multirow{2}{*}{FreeEOS\footnote{\citet{Irwin:2012uw}}} \\
			& $M < 0.7 M_\odot$: FreeEOS$^\text{f}$ & & $\log T < 7$: \citet{1990ApJ...350..300M} & \\ \hline
	He Fraction & $Y = 0.245 + 1.54 Z$ & $Y = 0.23 + 2.0 Z$ & $Y = 0.23 + 2.25 Z$ & $Y = 0.2485 + 1.78 Z$ \\
	$Z_\odot$ & 0.019 & 0.018 & 0.019 & 0.015 \\ 
	Solar Composition & \citet{1998SSRv...85..161G} & \citet{1993PhST...47..133G} & \citet{1993PhST...47..133G} & \citet{2011SoPh..268..255C} \\ \hline
	Atmospheres & PHOENIX\footnote{ \citet{2005ApJ...623..585F}} & \citet{1998AAS..130...65L} & ATLAS9\footnote{ \citet{2003IAUS..210P.A20C}} & ATLAS9 (Modified) \\ \hline
\end{tabular}
\end{table*}

In this work, four isochrone systems are considered: Dartmouth \citep{Dotter:2007fh}, Y$^2$ \citep{2001ApJS..136..417Y}, Padova \citep{2002AA...391..195G} and PARSEC \citep{2012MNRAS.427..127B}. Each model incorporates different physical assumptions (i.e. equation of state, radiative and conductive opacities), treatment of physical processes (i.e. convective transport, stellar atmospheres) and physical parameters (i.e. solar metallicity, initial He abundance, heavy-element mixture), all of which alter the resulting isochrone shape. Values for the input physics considered in this work are listed in table \ref{tab:isoparams}. Before comparing the isochrones to the observed data, some of the values in table \ref{tab:isoparams} can be standardized in order to simplify the final comparison.

\subsection{Age}
One of the main differences between isochrone systems used in this work is the range of available ages. M35 has a published age of 178 Myr \citep{2002AA...389..871D}, for which isochrones are available in the Padova and PARSEC systems. The nearest age in Y$^2$ is 200 Myr, while the youngest possible isochrone available for Dartmouth is 250 Myr. This work is interested in how the models treat main sequence stars, so this age difference is negligible; most stars on the main sequence will not have shifted in this 72 Myr span, given that all stars have finished their pre-main sequence evolution.

\subsection{Metallicity}
Many isochrone systems come pre-packaged in rough metallicity grids. For the isochrones to accurately match observed data, all must be interpolated to the metallicity of M35. Using previous WOCS work in \citet{2001ApJ...549..452B} the metallicity of M35 is [Fe/H]$=-0.21$, which was measured using high-resolution spectroscopy of 9 bright stars within the cluster. Interpolated Padova and PARSEC (v1.1) isochrones were pulled from the web via the interactive CMD 2.5 interface\footnote{\url{http://stev.oapd.inaf.it/cgi-bin/cmd}}. Y$^2$ isochrones include a FORTRAN routine that interpolates to a specified metallicity. Dartmouth isochrones utilize a similar interpolation web interface\footnote{\url{http://stellar.dartmouth.edu/\%7Emodels/webtools.html}} as Padova and PARSEC, however it is only available for ages $> 1$ Gyr, unsuitable for M35. Instead, another method must be employed for the Dartmouth isochrones.

\begin{figure} \centering
\includegraphics[trim = 0mm 248mm 20mm 20mm, clip, width=3.3in]{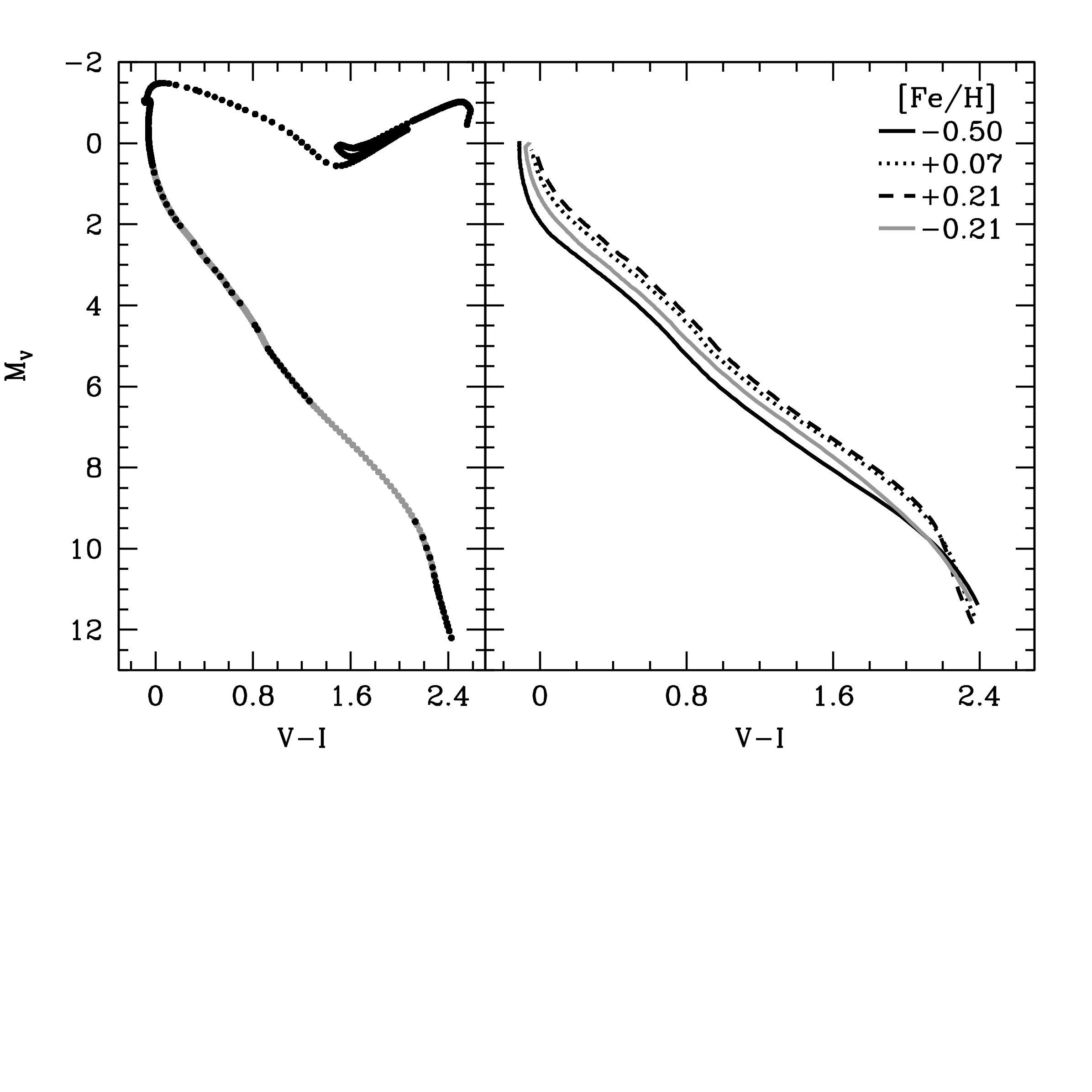}		
\caption{\emph{Left:} Interpolation onto a common mass grid for the [Fe/H]$=+0.07$ Dartmouth isochrone. Black dots are original isochrone grid points, grey are interpolated points in steps of 0.01 M$_\odot$. \emph{Right:} Interpolation in [Fe/H] of dartmouth isochrones to the M35 metallicity. Isochrones in grey are the three closest grid points used to create the interpolated isochrone. Note that interpolated isochrones only exist along the main sequence. All isochrones are for an age of 250 Myr.\label{fig:isointerp}}
\end{figure}

Starting with the Dartmouth system's pre-packaged metallicity grid, each isochrone is interpolated in mass, using a common spacing of 0.01 M$_\odot$. For each star in the new isochrone, stellar parameters ($\log g,L,T$) and magnitudes are cubically interpolated to the new mass value. This interpolation only works along the main sequence, where mass increases monotonically. This work is interested in how the isochrone treats the main sequence, and there is little evolution off of the main sequence in a young cluster like M35, so the loss of giants from the isochrone is acceptable. Once all isochrones are on a common mass grid, stellar parameters and magnitudes are quadratically interpolated in metallicity, using the nearest three isochrones in the grid to the desired [Fe/H]. For M35, [Fe/H]$=-0.50$, $+0.07$, $+0.21$ isochrones were used to interpolate the [Fe/H]$=-0.21$ isochrone. Figure \ref{fig:isointerp} shows the results of each of the interpolation steps.

\begin{figure*} \centering
\includegraphics[trim = 0mm 120mm 20mm 20mm, clip, width=6in]{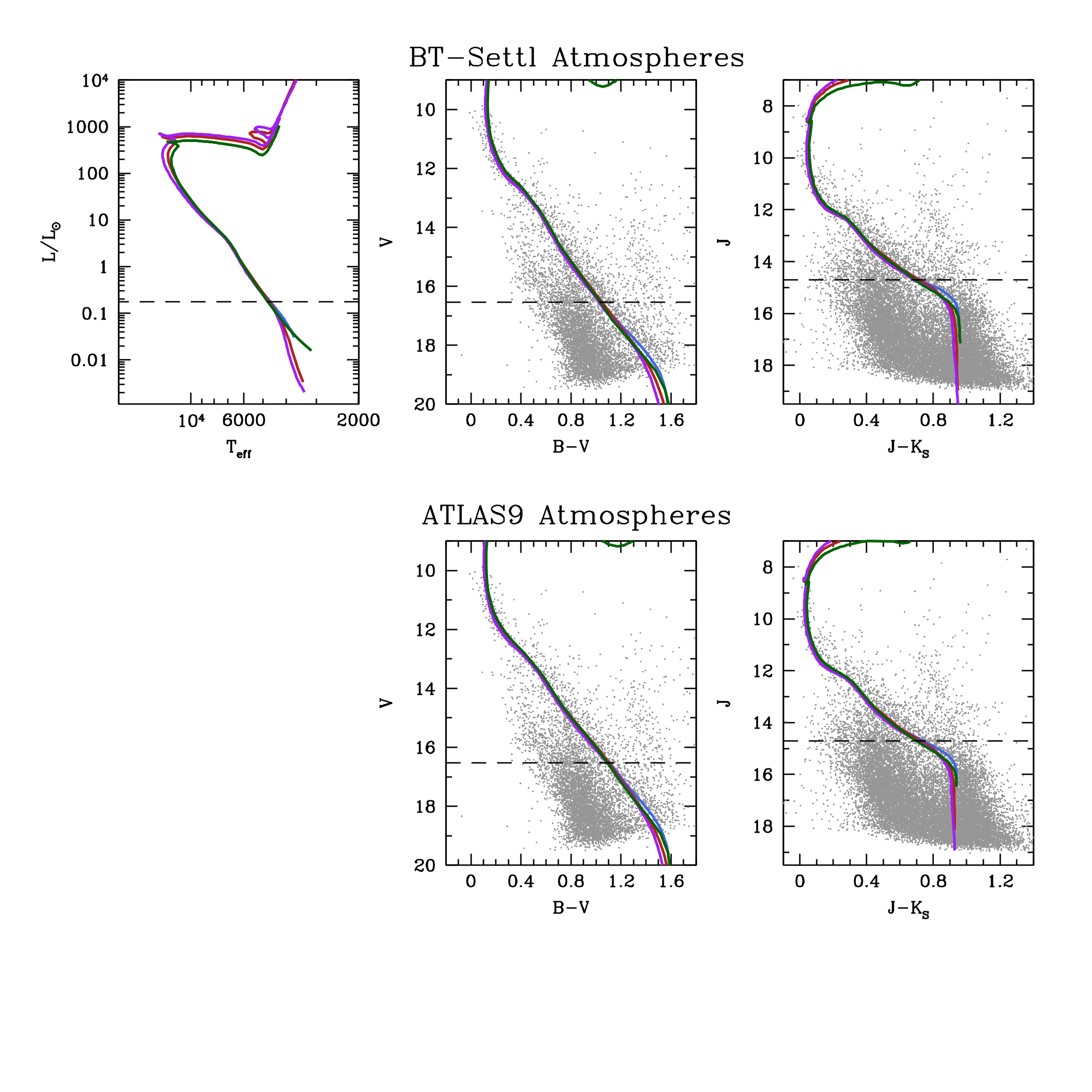}		
\caption{Comparison of the four isochrone systems and two atmosphere models used in this study. Grey dots are observed magnitudes from MOSAIC, NEWFIRM and 2MASS data described in section \ref{sec:data}. \emph{Blue:} 250 Myr Dartmouth isochrone. \emph{Red:} 178 Myr Padova isochrone. \emph{Purple:} 178 Myr PARSEC isochrone. \emph{Green:} 200 Myr Y$^2$ isochrone. All isochrones assume a distance of 870 pc, E($B-V$) = 0.22 and [Fe/H] = -0.21. Dashed lines indicate the position of the 0.7 M$_\odot$ model. \label{fig:isocompare}}
\end{figure*}

\subsection{Atmosphere Models}
Stars with similar internal parameters ($\log g, T$), but different color-temperature relations may look highly discrepant on a cluster CMD. Atmosphere models are standardized across all isochrone systems, allowing for a comparison of internal structure physics against observed data. Two atmosphere models are applied to the isochrones in this work: ATLAS9 \citep{2003IAUS..210P.A20C}, and BT-Settl \citep{2012RSPTA.370.2765A}.

ATLAS9 colors and bolometric corrections were downloaded from a pre-computed grid available online\footnote{\url{http://wwwuser.oat.ts.astro.it/castelli/colors.html}}. BT-Settl synthetic magnitudes, computed using the PHOENIX atmosphere code, were available for several solar abundances online\footnote{\url{http://phoenix.ens-lyon.fr/Grids/BT-Settl/}}. BT-Settl atmospheres, using solar abundances from \citet{2009ARA&A..47..481A}, were used in this work, differing from the ATLAS9 solar metallicity values of \citet{1998SSRv...85..161G}. The BT-Settl atmospheres carefully treat molecular absorption lines for cool stars, where ATLAS9 atmospheres are more incomplete.

Using the computed $\log g$ and $T_{\text{eff}}$ of each star in the isochrone, new magnitudes are computed using each of the atmosphere grids. ATLAS9 atmospheres are only available for temperatures greater than 3500K. Stars in the Dartmouth and Y$^2$ systems below this temperature were removed in the ATLAS9 isochrone.

\section{Analysis}

\begin{figure*} \centering
\includegraphics[trim = 0mm 20mm 20mm 0mm, clip, width=4.5in]{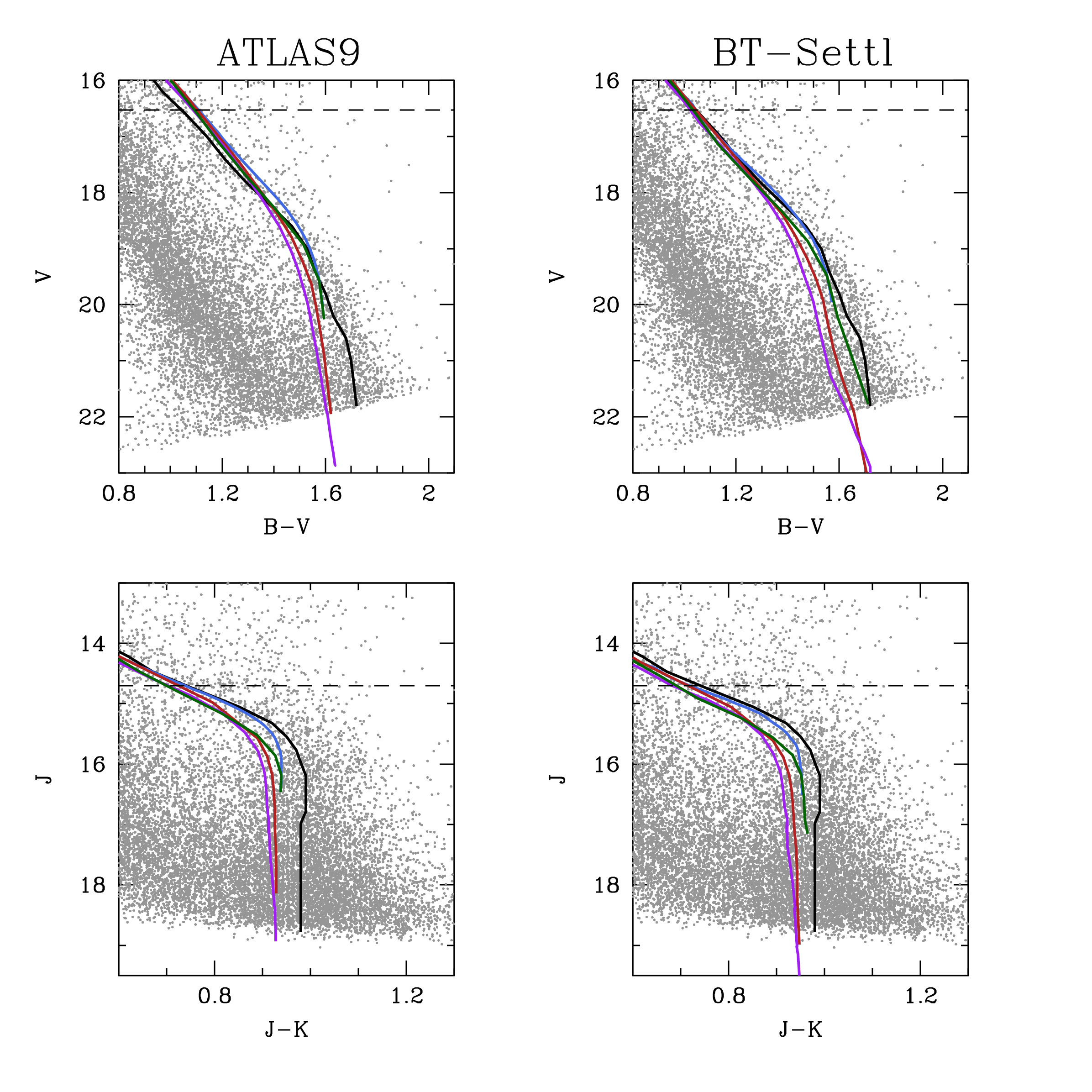}		
\caption{Zoomed-in CMDs of low mass portion of the main sequence. $J-K_S$ data from this work, $B-V$ data from \citet{2003AJ....126.1402K}. Black curves are hand-drawn empirical ridgelines. Dashed lines indicate the position of the 0.7 M$_\odot$ model. \label{fig:isocompare-zoomed}}
\end{figure*}

A comparison of the final isochrones to the observed data is shown in figure \ref{fig:isocompare}. There is little difference between the models for stars more massive than 0.7 M$_\odot$ ($V \sim 17$, $J \sim 15$ in M35), and all models match closely to data in this regime. For low mass stars, the isochrones begin to separate on the CMD. Unfortunately, our $UBVRI$ data is not deep enough to reach most of this region. Instead, $BV$ photometry of \citet{2003AJ....126.1402K} is used. A zoomed-in CMD of the low-mass regions of interest are shown in figure \ref{fig:isocompare-zoomed}.

Several conclusions about the input physics to the isochrone models can be made by comparison to the observed data.

\subsection{High-Mass Stars}
Each system employs vastly different physical assumptions for higher mass stars (M$>0.7$ M$_\odot$): for example, Dartmouth's equation of state model is a simple ideal gas, with a correction for coulomb interaction, while FreeEOS (used by PARSEC) handles ionization, degeneracy pressure and relativistic electron gas, yet the results end up nearly identical. The lack of difference is due to the fact that stars with masses larger than 0.7 M$_\odot$ have sufficiently low density that the addition of non-ideal effects in the equation of state does not produce appreciable shifts in the stellar model. Main sequence stellar parameters are insensitive to nearly all input parameters for main sequence stars with masses between 0.7 and 3.0 M$_\odot$.

\subsection{MHD Equation of State}
In the low-mass regime (M $<0.7$ M$_\odot$), the \citet{1990ApJ...350..300M} equation of state code (often referred to as MHD) begins to break down. Comparing Y$^2$ and Padova isochrones for low mass, there are many similarities in input physics, yet the resulting stellar parameters are quite different. The opacity codes and solar composition values are the same for both systems, and the solar metallicity and He abundances are very similar, yet the Padova isochrones predict much hotter temperatures than those from Y$^2$. The only difference between the two systems is the equation of state code.

Y$^2$'s OPAL EOS \citep{1996ApJ...456..902R} produces cooler low-mass stars, matching observation better than Padova. Padova isochrones' discrepant fits are not surprising, as the MHD EOS has been shown to produce inaccurate results even in the Sun \citep{1992AcA....42....5D}. The Padova set's results will all be affected due to the inaccurate EOS code.

\subsection{PARSEC Isochrones}
While the Padova isochrones' inconsistent temperatures can be explained by the MHD equation of state, their successor, the PARSEC system, predicts even hotter temperatures than those from Padova. These high temperatures produce colors which are the furthest from the observed main sequence of any of the sets examined in this work.

PARSEC's inconsistencies are not as easily explained as Padova's, using only the M35 data. There are many differences between PARSEC and the other systems, with PARSEC also being the only system with a different choice of solar metallicity. Comparing PARSEC isochrones to observations of clusters with different metallicity values will help determine whether the deviations are due to the metallicity adjustments, or any of the other differences in the PARSEC system.

\subsection{Atmospheres}
The switch from ATLAS9 to BT-Settl atmosphere models produces small but noticeable shifts in stellar magnitudes. Isochrones using the ATLAS9 atmosphere model are slightly offset from the observed main sequence, appearing too red above $V \sim 18$ and too blue below, as shown in figure \ref{fig:isocompare-zoomed}. Deviations from the observed main sequence are much smaller for the BT-Settl isochrones, with the Dartmouth and Y$^2$ isochrones matching closely to observation down to their faint limits, $V \sim 19$ and $V \sim 22$, respectively. The careful treatment of molecular lines in the BT-Settl atmospheres appears to enhance isochrone fits in the optical.

In the infrared, all isochrone and atmosphere combinations produce $J-K_S$ colors which are bluer than observed, as well as $J$ magnitudes which are fainter than the observed main sequence.  The treatment of molecular lines in the BT-Settl atmospheres yield accurate results in the optical, but may be incomplete in the IR. Further study of low-mass stellar atmospheres in the IR may reconcile this difference.

\section{Summary \& Future Work}

Several conclusions can be drawn from comparisons of theoretical isochrones to new photometry on the cluster M35:

\begin{itemize}
\item Theoretical magnitudes of stars with masses greater than 0.7 M$_\odot$ are insensitive to nearly all physical inputs. All isochrone and atmosphere systems produce accurate fits to the CMD for higher mass stars.
\item Low-mass stars (M $<$ 0.7 M$_\odot$) in the Padova system are hotter than observed. This is due to the inaccurate MHD Equation of State.
\item The careful treatment of molecular absorption lines in the BT-Settl atmosphere models yield better fits to the optical CMDs than ATLAS9, especially when applied to the Dartmouth or Y$^2$ systems.
\item All combinations of isochrone systems and atmosphere models yield bluer IR colors than observed data for stars with masses less than 0.7 M$_\odot$. BT-Settl atmospheres may lack the necessary molecular absorption information in the IR.
\end{itemize}

While this comparison has yielded several important insights into how various input physics alter the fit of an isochrone, only so much can be determined from a single cluster. A future paper will compare these same isochrone systems to open clusters of varying ages and metallicities in order to further improve these conclusions.

\section{Acknowledgements}

The authors would like to acknowledge graduate thesis support from NOAO as well as financial support from the Texas Space Grant Consortium. This research uses services or data provided by the NOAO Science Archive. NOAO is operated by the Association of Universities for Research in Astronomy (AURA), Inc. under a cooperative agreement with the National Science Foundation.

\bibliographystyle{apj}
\bibliography{PaperI}

\end{document}